\documentclass[12pt]{article}
\usepackage{amssymb,amsfonts}
\usepackage{graphics,psboxit,amsmath}
\usepackage{subfigure}
\usepackage{graphicx}
\usepackage{verbatim}
\usepackage{color}
\usepackage{hyperref}
\hypersetup{colorlinks}

\definecolor{darkred}{rgb}{1,0,0}
\definecolor{darkgreen}{rgb}{0,0.5,0}
\definecolor{darkblue}{rgb}{0,0,1}
\definecolor{orange}{rgb}{1,0.5,0}
\definecolor{green}{rgb}{0,1,0}
\definecolor{purple}{rgb}{.5,0,1}

\hypersetup{ colorlinks,
linkcolor=darkred,
filecolor=darkgreen,
urlcolor=darkblue,
citecolor=darkblue,
linktocpage=true }

\definecolor{markcolor}{rgb}{.25,0,1}

\definecolor{markcolor2}{rgb}{1,0,0}

\definecolor{markcolor3}{rgb}{0,1,0}

\newcommand{\poisbrac}[2]{\Big\{#1\hspace{2.0pt}
 \mbox{\raisebox{2pt}{$\otimes$}}\hspace{-5.85pt}
 \mbox{\raisebox{-3pt}{,}}\hspace{5pt}#2\Big\}}

\def\hybrid{\topmargin -10pt    \oddsidemargin 0pt %%%%%%%%%%%%%% Archive-30pt
        \headheight 0pt \headsep 0pt
        \textwidth 16.5cm      % A4 paper
        \textheight 23cm       % A4 paper
        \marginparwidth .875in
        \parskip 5pt plus 1pt   \jot = 1.5ex}

\hybrid

\catcode`\@=11

\def\marginnote#1{}
\newcount\hour
\newcount\minute
\newtoks\amorpm
\hour=\time\divide\hour by60 \minute=\time{\multiply\hour by60
\global\advance\minute by-\hour}
\edef\standardtime{{\ifnum\hour<12 \global\amorpm={am}%
        \else\global\amorpm={pm}\advance\hour by-12 \fi
        \ifnum\hour=0 \hour=12 \fi
        \number\hour:\ifnum\minute<10 0\fi\number\minute\the\amorpm}}
\edef\militarytime{\number\hour:\ifnum\minute<10 0\fi\number\minute}
\def\draftlabel#1{{\@bsphack\if@filesw {\let\thepage\relax
   \xdef\@gtempa{\write\@auxout{\string
      \newlabel{#1}{{\@currentlabel}{\thepage}}}}}\@gtempa
   \if@nobreak \ifvmode\nobreak\fi\fi\fi\@esphack}
        \gdef\@eqnlabel{#1}}
\def\@eqnlabel{}
\def\@vacuum{}
\def\draftmarginnote#1{\marginpar{\raggedright\scriptsize\tt#1}}

\def\draft{\oddsidemargin -.5truein
        \def\@oddfoot{\sl preliminary draft \hfil
        \rm\thepage\hfil\sl\today\quad\militarytime}
        \let\@evenfoot\@oddfoot \overfullrule 3pt
        \let\label=\draftlabel
        \let\marginnote=\draftmarginnote
   \def\@eqnnum{(\theequation)\rlap{\kern\marginparsep\tt\@eqnlabel}%
\global\let\@eqnlabel\@vacuum}  }

\def\draft2{
        \def\@oddfoot{\sl preliminary draft \hfil
        \rm\thepage\hfil\sl\today\quad\militarytime}
        \let\@evenfoot\@oddfoot \overfullrule 3pt
        \let\label=\draftlabel
        \let\marginnote=\draftmarginnote
   \def\@eqnnum{(\theequation)\rlap{\kern\marginparsep\tt\@eqnlabel}%
\global\let\@eqnlabel\@vacuum}  }

\def\preprint{\twocolumn\sloppy\flushbottom\parindent 2em
        \leftmargini 2em\leftmarginv .5em\leftmarginvi .5em
        \oddsidemargin -.5in    \evensidemargin -.5in
        \columnsep .4in \footheight 0pt
        \textwidth 10.in        \topmargin  -.4in
        \headheight 12pt \topskip .4in
        \textheight 6.9in \footskip 0pt
        \def\@oddhead{\thepage\hfil\addtocounter{page}{1}\thepage}
        \let\@evenhead\@oddhead \def\@oddfoot{} \def\@evenfoot{} }

\def\numberbysection{\@addtoreset{equation}{section}
        \def\theequation{\thesection.\arabic{equation}}}

\def\underline#1{\relax\ifmmode\@@underline#1\else
        $\@@underline{\hbox{#1}}$\relax\fi}

\def\titlepage{\@restonecolfalse\if@twocolumn\@restonecoltrue\onecolumn
     \else \newpage \fi \thispagestyle{empty}\c@page\z@
        \def\thefootnote{\fnsymbol{footnote}} }

\def\endtitlepage{\if@restonecol\twocolumn \else \newpage \fi
        \def\thefootnote{\arabic{footnote}}
        \setcounter{footnote}{0}}  %\c@footnote\z@ }

\catcode`@=12 \relax

\def\figcap{\section*{Figure Captions\markboth
        {FIGURECAPTIONS}{FIGURECAPTIONS}}\list
        {Figure \arabic{enumi}:\hfill}{\settowidth\labelwidth{Figure
999:}
        \leftmargin\labelwidth
        \advance\leftmargin\labelsep\usecounter{enumi}}}
 \relax
\def\tablecap{\section*{Table Captions\markboth
        {TABLECAPTIONS}{TABLECAPTIONS}}\list
        {Table \arabic{enumi}:\hfill}{\settowidth\labelwidth{Table
999:}
        \leftmargin\labelwidth
        \advance\leftmargin\labelsep\usecounter{enumi}}}
 \relax
\def\reflist{\section*{References\markboth
        {REFLIST}{REFLIST}}\list
        {[\arabic{enumi}]\hfill}{\settowidth\labelwidth{[999]}
        \leftmargin\labelwidth
        \advance\leftmargin\labelsep\usecounter{enumi}}}
 \relax
\makeatletter
\newcounter{pubctr}
\def\publist{\@ifnextchar[{\@publist}{\@@publist}}
\def\@publist[#1]{\list
        {[\arabic{pubctr}]\hfill}{\settowidth\labelwidth{[999]}
        \leftmargin\labelwidth
        \advance\leftmargin\labelsep
        \@nmbrlisttrue\def\@listctr{pubctr}
        \setcounter{pubctr}{#1}\addtocounter{pubctr}{-1}}}
\def\@@publist{\list
        {[\arabic{pubctr}]\hfill}{\settowidth\labelwidth{[999]}
        \leftmargin\labelwidth
        \advance\leftmargin\labelsep
        \@nmbrlisttrue\def\@listctr{pubctr}}}
 \relax
\makeatother

\def\be{\begin{equation}}
\def\ee{\end{equation}}
\def\ba{\begin{eqnarray}}
\def\ea{\end{eqnarray}}

\def\del{\partial}

\newcommand{\fr}[1]{\mathfrak{#1}}

\def\g{\gamma}

\def\d{\delta}

\def\m{\mu}

\def\Om{\Omega}
\def\l{\lambda}

\def\s{\sigma}

\def\no{\noindent}

\def\qq{\qquad}

\def\IR{\relax{\rm I\kern-.18em R}}

\def\bse{\begin{small}\begin{equation*}}
\def\ese{\end{equation*}\end{small}}

\begin{document}
%\draft2

%\renewcommand{\theequation}{\arabic{equation}}
%\renewcommand{\theequation}{\thesection.\arabic{equation}}

\renewcommand{\theequation}{\thesection.\arabic{equation}}
\csname @addtoreset\endcsname{equation}{section}

\newcommand{\eqn}[1]{(\ref{#1})}

\begin{titlepage}
\begin{center}
\strut\hfill
\vskip 1.3cm

%\hfill  [hep-th]\\

\vskip .5in

{\Large \bf Sigma models in the presence of dynamical\\ point-like defects}

\vskip 0.5in

{\bf Anastasia Doikou and Nikos Karaiskos} \vskip 0.1in

{\footnotesize Department of Engineering Sciences, University of Patras,\\
GR-26500 Patras, Greece}

\vskip .1in

%\vskip -.15in

{\footnotesize {\tt E-mail:$\{$adoikou, nkaraiskos$\}@$upatras.gr}}\\

\end{center}

\vskip .6in

\centerline{\bf Abstract}
Point-like Liouville integrable dynamical defects are introduced in the context of the Landau-Lifshitz
and Principal Chiral (Faddeev-Reshetikhin) models. Based primarily on the underlying quadratic algebra we identify the
first local integrals of motion, the associated Lax pairs as well as the relevant sewing conditions around the defect point.
The involution of the integrals of motion is shown taking into account the sewing conditions.

\no

\vfill

\end{titlepage}
\vfill \eject

%\def\baselinestretch{1.2}
%\baselineskip 10 pt
%\noindent

\tableofcontents

\section{Introduction}

The issue of integrable defects has been quite intriguing, and a considerable number
of studies have been devoted to this particular problem, both at classical and
quantum level \cite{fitsvewi}--\cite{avan-doikou-defect2}. From a physical point of view, the insertion
of defects or impurities within a given physical system renders the latter more
interesting and realistic. Applications of such studies are important for a number of
different disciplines, including models in condensed matter theory
(see e.g. \cite{fitsvewi, Frahm, Affleck2}) and
quantum information (see e.g. \cite{Osenda, Santos}). It is desirable then to introduce defects in integrable
theories, in such a way that integrability is preserved, so that the corresponding tools
can be used in order to obtain precise information about the defect model.

A systematic algebraic formulation for describing Liouville integrable point-like
defects was recently introduced in \cite{avan-doikou-defect1, avan-doikou-defect2}.
The description was primarily based on the underlying quadratic algebra satisfied by
the bulk monodromy matrices, which describe the left and right theories, as well as
by the defect L-matrix. In fact, this is the main necessary requirement so that
Liouville integrability may be by construction guaranteed. The main steps of the
proposed algebraic methodology are provided in the subsequent sections, however
for a more detailed description of the process we refer the interested reader to
\cite{avan-doikou-defect1, avan-doikou-defect2}.

An efficient description of defects becomes more intricate
at the level of classical integrable field theories, where usually the defect is
introduced as a discontinuity ({\it jump}) together with suitable sewing conditions
\cite{BCZ1, BCZ2}. In the present scheme the sewing conditions naturally emerge as
continuity conditions on the time components of the Lax pair around the defect
point. It was also rigorously proven in \cite{avan-doikou-defect1} that the sewing
conditions are compatible with the hierarchy of the Hamiltonians, a fact that ensures
that the proposed formulation is well defined and consistent.

In the present work,
we consider dynamical point-like defects in the context of sigma models, such as the
Landau-Lifshitz (L-L) model and a variation of the familiar principal chiral model (PCM),
the so called Faddeev-Reshetikhin (F-R) \cite{Faddeev:1985qu} model, in such a way that
integrability is preserved. It is worth noting that this kind of systems, in addition
to their own physical and mathematical value have also attracted considerable interest
lately due to the fact that they typically arise within the AdS/CFT context (see e.g.
\cite{kaza, hern, Avan:2010xd} and references therein). Thus, it would be of great
significance to investigate possible relevant physical implications in this particular
frame. Recall also that the typical PCM possesses a non-ultra local algebra rendering
its quantization a very intricate task, whereas its variation, the F-R model is associated
to a more familiar ultra-local algebra, much easier to deal with. In any case, both
PCM and F-R models share the same Lax pair, and consequently the same equations of motion,
hence they are physically quite similar. Therefore, the findings presented for the F-R
model are naturally relevant for the conventional PCM model.

Based on the formulation of \cite{avan-doikou-defect1, avan-doikou-defect2} we introduce
the defect matrix, and the associated modified monodromy matrix. Using this as our
starting point we extract the first couple of the local integrals of motion, and the
corresponding time components of the Lax pairs for both models. Due to analyticity
requirements imposed on the time components of the Lax pairs, certain sewing conditions
emerge at the defect point. The involution of the charges is shown based on the
underlying algebra, and the invariance of the sewing conditions under the Hamiltonian
action is also explicitly checked.

\section{The Landau-Lifshitz model with defect}
We first consider integrable point-like  dynamical defects in the context of the isotropic
Landau-Lifshitz model. We assume periodic boundary
conditions and restrict our attention to the $\mathfrak{su}_2$ classical algebra \cite{Faddeev:1987ph}.
Results regarding higher rank algebras may also be considered \cite{Doikou:2011rf}.

The equations of motion associated with the isotropic Landau-Lifshitz model are given as:
\be
\frac{\del \vec{S}}{\del t}=i \vec{S} \times \frac{\partial^2 \vec{S}}{\partial x^2}\, ,
\label{e.o.m.}
\ee
where the vector-valued functions $\vec{S}(x)= (S_1(x),S_2(x),S_3(x))$ which describe
the physical quantities of the model take values on the unit 2-sphere, i.e.
 $\vec{S}\cdot \vec{S}=1$, and
satisfy an $\fr{su}_2$ Poisson structure given by the following Poisson brackets
\be
\Big \{ S_a(x),\ S_b(y)\Big \} = 2i\varepsilon_{abc}\, S_c(x)\, \d(x-y)\, ,
\ee
with $\varepsilon_{abc}$ being the totally antisymmetric Levi-Civita tensor with
value $\varepsilon_{123} = 1$. We note that throughout the present section we shall
also use the following linear combinations of the fields $S_i(x)$:
\be
S^{\pm}(x) = \frac{1}{2}\big(S_1 (x) \pm i S_2(x)\big)\, .
\ee
The Hamiltonian and the momentum of the model are given by the following expressions
\ba
\mathcal{H} & = & - \frac{1}{4}\int_{-L}^L dx \,\Big(\frac{\del \vec{S}}{\del x}\Big)^2\, ,\cr
\mathcal{P} & = &  {i\over 2} \int_{-L}^{L} dx ~\frac{S_1 S_2' - S_1' S_2 }{1+S_3}\, ,
\ea
where the prime denotes derivative with respect to $x$. Using the Hamiltonian, the
equations of motion can also be expressed as
\be
\frac{\del \vec{S}}{\del t} = \Big \{\mathcal{H},~\vec{S}\Big \}\, .
\label{emh}
\ee

\subsection{The Lax and defect operators}
We may now  introduce a single point-like defect of  dynamical type in the L-L model at the point $x_0$. The
starting point for analyzing the system is the derivation of the modified monodromy matrix,
which is built as a co-action \cite{avan-doikou-defect1}
\ba
\mathcal{T}(L,-L,\l) & = & T^+(L, x_0, \l)\, \mathcal{D}(x_0,\l)\, T^-(x_0,-L,\l) \cr
& = & \mathcal{P}\textrm{exp}\Big(\int_{x_0^+}^L dx~\mathbb{U}^+(x)\Big) \,
\mathcal{D}(x_0,\l)
\, \mathcal{P}\textrm{exp}\Big(\int_{-L}^{x_0^-} dx~\mathbb{U}^-(x)\Big)~.
\ea
Recall the Lax pair $(\mathbb{U},\mathbb{V})$ for the bulk L-L model
\be
\mathbb{U}(x) =  \frac{1}{\l}
\begin{pmatrix}
 \frac{S_3}{2} & S^- \cr
 S^+ & -\frac{S_3}{2}
\end{pmatrix}
\equiv \frac{1}{2\l}\mathcal{S}, \qquad
\mathbb{V}(x) =  \frac{1}{2  \l^2} \mathcal{S} - \frac{1}{2\l} \frac{\del\mathcal{S}}{\del x}
\mathcal{S}\, .
\label{U_V_bulk}
\ee
As indicated in e.g. \cite{avan-doikou-defect1}, the defect operator is required to
satisfy the same quadratic algebra as the monodromy matrix, i.e.
\be
\poisbrac{\mathcal{D}(\l)}{\mathcal{D}(\m)}
= \Big [r(\l-\m),\, \mathcal{D}(\l) \otimes \mathcal{D}(\m)\Big ]\, .
\label{defect_quad}
\ee
In the case of the L-L model, the classical $r$-matrix is the Yangian solution \cite{yang}
\be
r(\l) = - \frac{\mathcal{P}}{\l}\, ,
\ee
with $\mathcal{P}$ being the permutation operator: $\mathcal{P}(\vec{a}\otimes \vec{b}) =
\vec{b}\otimes \vec{a}$. It turns out that the generic defect operator satisfying the
quadratic Poisson structure with the Yangian $r$-matrix has the following form
\be
\mathcal{D}(\l) = \l ~\mathbb{I} +
\begin{pmatrix}
  \xi^z & \xi^- \cr \xi^+ & -\xi^z
\end{pmatrix}, \label{defectmatrix}
\ee
due to \eqn{defect_quad} the elements satisfy the $\fr{sl}_2$ algebraic relations
\ba
&& \Big \{\xi^z ,~ \xi^{\pm} \Big \} = \pm \xi^{\pm} \cr
&& \Big \{\xi^+ ,~ \xi^- \Big \} = 2\xi^z. \label{sl}
\ea
Notice that the associated Casimir is given as: ${\cal C} = (\xi^z)^2 + \xi^+ \xi^-$ hence,
the algebra (\ref{sl}) may be parameterized by two free fields. Similarly for (\ref{U_V_bulk}).

The continuum ``bulk'' monodromy matrices $T^{\pm}$ satisfy the usual differential
equation
\be
\frac{\del T^{\pm}(x,y; \l)}{\del x} = {\mathbb U}^{\pm}(x,\lambda) T^{\pm}(x,y; \lambda)~,
\label{dif1}
\ee
and the zero curvature condition is then expressed as:
\be
\dot{\mathbb U}^{\pm}(x,t) - {\mathbb V}^{\pm '}(x,t) + \Big [{\mathbb U}^{\pm}(x, t),
{\mathbb V}^{\pm}(x, t) \Big ] =0\, , ~~~~~x \neq x_0~.
\label{zero1}\ee
On the defect point in particular the zero curvature condition is formulated as
(see e.g. \cite{doikou-defect, avan-doikou-defect1} for more details)
\be
\frac{d}{dt}\,  \mathcal{D}(x_0)= \widetilde {\mathbb V}^+(x_0) \mathcal{D}(x_0)
- \mathcal{D}(x_0) \widetilde {\mathbb V}^{-}(x_0)\, ,
\label{zerod}
\ee
and describes explicitly the {\it jump} occurring across the defect point. This will be a
major consistency check of the prescription followed here. The time components
$\widetilde {\mathbb V}^{\pm}$ of the Lax pairs for the left and right bulk theories
as well as the defect point will be explicitly derived subsequently.

\subsection{Local integrals of motion}
We first derive the tower of involutive local integrals of motion. Regarding
the left and right bulk parts, we consider the usual ansatz for the respective monodromy matrices
\be
T^{\pm}(x,y;\lambda) = (1+W^{\pm}(x))~e^{Z^{\pm}(x,y)}~(1+W^{\pm}(y))^{-1} ~,
\label{transatz}
\ee
with $W^{\pm}$ and $Z^{\pm}$ being purely off-diagonal and diagonal matrices respectively. We also
assume that they admit an expansion in terms of the spectral parameter $\l$ as
\be
W^{\pm}(x,\l)=\sum_{n=0}^{\infty}\l^n W^{\pm}_n(x)~, \qq Z^{\pm}(x,y,\l)=\sum_{n=-1}^{\infty}\l^n Z^{\pm}_n(x,y)~.
\label{WZexp}
\ee
Substituting the ansatz \eqn{transatz} into the relation \eqn{dif1}, and splitting the resulting equation
into a diagonal and an off-diagonal part one obtains
\ba
&& \frac{d W^{\pm}}{dx} + W^{\pm} \mathbb{U}_d^\pm - \mathbb{U}_d^\pm W^{\pm} +  W^{\pm}
\mathbb{U}_a^{\pm} W^{\pm}- \mathbb{U}^{\pm}_a =0,\cr
&& \frac{\del Z^{\pm}}{\del x} = \mathbb{U}_d^\pm + \mathbb{U}_a^{\pm} W^{\pm}~.
\label{split0}
\ea
The off-diagonal part corresponds to a typical Riccati type differential equation.
Solving this set of differential equations provides explicit expressions of the
$W^{\pm},Z^{\pm}$ matrices. As for the local integrals of motion, recall that they are
provided by the generating functional
\be
\mathcal{G}(\l) = \ln \Big(\textrm{tr}\, \mathcal{T}(\l) \Big)~.
\ee
Substituting the ansatz \eqn{transatz}, the latter expression can be formulated as:
\be
{\cal G}(\lambda) = \ln \mathrm{tr} \Big [e^{Z^+(L, x_0)} (1+W^+(x_0))^{-1}
~\mathcal{D}(x_0) ~ (1+W^-(x_0)) e^{Z^-(x_0, -L))}  \Big ]~,
\ee
where Schwartz boundary conditions at the endpoints $x=\pm L$ have been implemented.

Solving \eqn{split0} one determines $W^{\pm}$ and $Z^{\pm}$ order by order. The first
three terms of the expansion suffice to compute the Hamiltonian and the momentum of the
bulk theories, and are found to be
\ba
&& \mathcal{O}(1/\l): \qquad
W_0^{\pm} = \begin{pmatrix}
       0 & -\bar{a} \cr a & 0
      \end{pmatrix}, \qquad a = \frac{1-S_3}{2S^-} = \frac{2 S^+}{1+S_3},\cr
&& \mathcal{O}(\l^{0}): \qquad
W_1^{\pm} = \begin{pmatrix}
       0 & -\bar{a}' \cr -a' & 0
      \end{pmatrix},\cr
&& \mathcal{O}(\l): \qquad
W_2^{\pm} = \begin{pmatrix}
       0 & -\bar{a}''+(\bar{a}')^2S^+ \cr a''-(a')^2S^- & 0
      \end{pmatrix}~.
\label{explicitW}
\ea
Note that $\bar{a}$ is obtained by interchanging
$S^- \leftrightarrow S^+$. Correspondingly, we find that $Z^{\pm}$ is
given by
\be
Z^{\pm}(x,y,\l)=\frac{1}{2\l}(x-y)\, \s^z + \sum_{n=1}^{\infty}\l^{n-1}
\int_y^x  dz ~(S^+\s^-+S^-\s^+)W_n^{\pm}(z)~.
\ee
At order $\mathcal{O}(1/\l)$ one finds
\be
Z_{-1}^{\pm}=\frac{1}{2}(x-y)\, \s^z~.
\ee
This term is important, since it provides the leading contribution of $e^Z$ as $\l\to0$,
a fact that is used when expanding the generating function $\mathcal{G}$.
The next two orders provide the first two physical integrals of motion for the right and
left theories. More specifically, one concludes that
\be
(Z_0^{\pm})_{11} = - \mathcal{P}^{\pm}, \qquad \textrm{and} \qquad
(Z_1^{\pm})_{11} = \mathcal{H}^{\pm}~.
\ee
It should be stressed out that the implementation of special boundary conditions at the
defect point $x_0$ amounts to the emergence of certain boundary type terms in both charges,
which have to be taken into account.

Expanding then the generating function ${\cal G}$ in powers of $\l$ leads to the local integrals
of motion of the defect L-L model. More precisely, the generating function of the
local integrals of motion becomes
\be
{\cal G}(\lambda) =Z_{11}^{+}(\lambda) + Z^{-}_{11}(\lambda) +
\ln \Big[\big(1+W^+(x_0)\big)^{-1} \, \mathcal{D}(x_0) \, \big(1+W^-(x_0)\big)\Big]_{11}\, ,
\label{genfundef}
\ee
so that the terms $Z_{11}^{\pm}$ provide the left and right bulk charges computed above,
whereas the third term gives the defect contribution.

Gathering all the information given above we derive the explicit expressions for the
momentum and Hamiltonian of the system:
\ba
{\cal  P} &=& {i\over 2}\int_{-L}^{x_0} dx\ {S_1 S_2' - S_2 S_1' \over 1 +S_3} + {i\over 2}\int_{x_0}^{L} dx\ {\tilde S_1 \tilde S_2' - \tilde S_2 \tilde S_1' \over 1 + \tilde S_3} \cr &+&{1\over 2} \ln(1 + \tilde S_3(x_0))-{1\over 2} \ln (1 + S_3(x_0)) -\ln {\cal D}^{(0)}(x_0)\, , \cr
% {\cal H} &=& -{1 \over 4} \int_{-L}^{x_0} \left ( \Big({\partial S_1 \over \partial x}\Big )^2 + \Big ({\partial S_2 \over \partial x}\Big )^2 + \Big ({\partial S_3 \over \partial x}\Big )^2\right)-{1 \over 4} \int_{x_0}^{L} \left ( \Big({\partial \tilde S_1 \over \partial x}\Big )^2 + \Big ({\partial \tilde S_2 \over \partial x}\Big )^2 + \Big ({\partial \tilde S_3 \over \partial x}\Big )^2\right)  \cr
{\cal H} &=& -{1 \over 4} \int_{-L}^{x_0} \Big ( {\partial \vec{S} \over \partial x}
 \Big)^2-{1 \over 4} \int_{x_0}^{L} \Big ( {\partial \tilde{\vec{S}} \over \partial x}
\Big)^2   +{\cal D}^{(1)}(x_0) \cr
&-&{1\over 2} \big(\tilde S_1(x_0) - \tilde S_2 (x_0)\big)\, a^{+'}(x_0)
+ {1\over 2}\big(S_1(x_0) - S_2 (x_0)\big)\, a^{-'}(x_0)\, ,
\ea
where we define
\be
a^+ =\frac{1-\tilde S_3}{2\tilde S^-} = \frac{2 \tilde S^+}{1+\tilde S_3}, ~~~~~a^- =\frac{1-S_3}{2S^-} = \frac{2 S^+}{1+S_3} \, ,
\ee
and $\bar a^{\pm}$ are defined analogously.
Clearly the variables $S_i$ describe the left bulk theory, whereas the variables $\tilde S_i$ describe the right bulk theory.
Note also the emergence of some non-trivial boundary type terms at the defect point, in
both charges. Regarding the defect contributions, we have found the following
explicit expressions for the first two physical charges
\ba
\mathcal{D}^{(0)} & = & \frac{1}{\fr{d^+}}
\Big[ a^-\xi^- + \bar{a}^+ \xi^+ + (1 - \bar{a}^+ a^-)\xi^z\Big] \cr
\mathcal{D}^{(1)} & = & \frac{1}{\fr{J}} \left[ \fr{d}^+(1+a^-\bar{a}^+)
+ \left(\bar{a}^{+'} + (\bar{a}^+)^2 a^{+'}\right)\xi^+  \right.\cr
&& - \left( (\bar{a}^{+'}a^+ - \bar{a}^+ a^{+'})a^- + \fr{d}^+ a^{-'} \right)\xi^- \cr
&  &  - \left.\left(
\bar{a}^{+'} (a^+ + a^-) + \bar{a}^+ (\bar{a}^+ a^{+'} a^- -a^{+'}  -\fr{d}^+ a^{-'}) \right)\xi^z \right]\, ,
\label{deffstch}
\ea
with
\be
 \fr{J} \equiv \fr{d}^+(a^-\xi^- + \bar{a}^+ \xi^+ + (1 - \bar{a}^+ a^-)\xi^z)
= (\fr{d}^+)^2 \mathcal{D}^{(0)}, \qquad \fr{d}^\pm \equiv 1 + a^\pm \bar{a}^\pm \, .
\label{d_defin}
\ee
These two terms are the defect contributions in the momentum and Hamiltonian charges
respectively. For the sake of clarity, we regard from now on $a^{\pm},\bar{a}^{\pm}$ as the fundamental
variables, instead of the fields $S_i,~\tilde S_i$. We should also note that these expressions
simplify, when one considers the sewing conditions on the defect point, as will be transparent
 in the subsequent section.

\subsection{The modified Lax pair}
From the general theory of classical integrable systems, the  time components $\mathbb{V}$
of the Lax pair are given by \cite{avan-doikou-defect1, Faddeev:1987ph}
\be
\mathbb{V}(x,\l,\m) = t^{-1}(\l)\, \textrm{tr}_a\Big(\mathcal{T}_a(L,x,\l)\, r_{ab}(\l,\m) \,
\mathcal{T}_a(x,-L,\l)\Big)~.
\ee
In the case where defects are taken into account, one has to compute $\mathbb{V}$ in
the bulk, as well as at the defect point. If the $r$-matrix of
the model is the Yangian, as it happens for the L-L model, the corresponding time components
are simplified and for a single point-like defect are expressed as \cite{avan-doikou-defect1}
\ba
&& \mathbb{V}^+(x,\l,\m) = \frac{t^{-1}}{\l-\m}T^+(x,x_0) \mathcal{D}(x_0)
T^-(x_0,-L)T^+(L,x) \cr
&& \mathbb{V}^-(x,\l,\m) = \frac{t^{-1}}{\l-\m}T^-(x,-L)T^+(L,x_0)\mathcal{D}(x_0)
T^-(x_0,x)\cr
&& \mathbb{\widetilde{V}}^+(x_0,\l,\m) = \frac{t^{-1}}{\l-\m} \mathcal{D}(x_0)T^-(x_0,-L)
T^+(L,x_0) \cr
&& \mathbb{\widetilde{V}}^-(x_0,\l,\m) = \frac{t^{-1}}{\l-\m} T^-(x_0,-L)T^+(L,x_0)
\mathcal{D}(x_0)~.
\ea
We provide explicit expressions for the first three orders of the $\mathbb{V}$
operators below. It should be clear that the following expressions hold for the left bulk
part
\ba
\mathcal{O}(\l^0): &&  \mathbb{V}_{(0)}^-(x,\m) = -\frac{\mathbb{I} + \mathcal{S}}{2\m}
= -\frac{1}{2\m}
  \begin{pmatrix}
   1 + S_3 & 2S^-  \cr
   2S^+  & 1- S_3
  \end{pmatrix}\\
\mathcal{O}(\l): &&  \mathbb{V}_{(1)}^-(x,\m) = \frac{1}{\m}
\begin{pmatrix}
  S^+S^{-'} - S^-S^{+'} & S^-S_3' - S_3 S^{-'}	\cr
  S_3 S^{+'} - S_3' S^+ &  S^-S^{+'} - S^+S^{-'}
\end{pmatrix}
+\frac{1}{\m}\mathbb{V}_{(0)}^-(x,\m).\nonumber
\ea
Similar expressions hold for ${\mathbb V}_{(i)}^+$, but with $S^i \to \tilde S^i$.
At order $\mathcal{O}(\l)$ one naturally finds the time component associated with the
Hamiltonian. For the sake of clarity, as noted before, it is more useful to
write down the operators in terms of the fields $a^{\pm}$, defined in
\eqn{explicitW}
\ba
\mathcal{O}(\l^0):  && \mathbb{V}_{(0)}^\pm(x,\m) = -\frac{1}{\m}
\frac{1}{\fr{d}^\pm}
  \begin{pmatrix}
   1  & \bar{a}^{\pm}  \\
   a^{\pm}  & a^{\pm}\bar{a}^{\pm}
  \end{pmatrix}\\
\mathcal{O}(\l):  && \mathbb{V}_{(1)}^\pm(x,\m) = \frac{1}{\m^2}
\frac{1}{(\fr{d}^\pm)^2} \ \begin{pmatrix}
   \bar{a}^{\pm'}(a^\pm - \bar{a}^{\pm})  - \fr{d}^\pm
 & -\bar{a}^{\pm'} -\bar{a}^{\pm}(\fr{d}^\pm +\bar{a}^{\pm} a^{\pm'})	\cr
  a^{\pm}(a^{\pm}\bar{a}^{\pm'}- \fr{d}^\pm)  + a^{\pm'}
& a^{\pm'}(\bar{a}^{\pm}-a^{\pm}) -a^{\pm}\bar{a}^{\pm} \fr{d}^\pm
\end{pmatrix}\, .\nonumber
\ea
At the defect point, we have the following expressions at the first order
\ba
\mathbb{\widetilde{V}}_{(0)}^+(x_0,\m) & = &-\frac{1}{\m}\frac{\fr{d}^+}{\fr{J}}
\begin{pmatrix}
 a^- \xi^- + \xi^z  & \bar{a}^+(a^-\xi^- + \xi^z) \cr
 \xi^+ - a^- \xi^z & \bar{a}^+(\xi^+ - a^- \xi^z)
\end{pmatrix} \cr
\mathbb{\widetilde{V}}_{(0)}^-(x_0,\m) & = & - \frac{1}{\m}\frac{\fr{d}^+}{\fr{J}}
\begin{pmatrix}
 \bar{a}^+ \xi^+ + \xi^z  &  \xi^- - \bar{a}^+ \xi^z \cr
 a^-(\bar{a}^+\xi^+ + \xi^z) & a^-(\xi^- - \bar{a}^+ \xi^z)
\end{pmatrix}~.
\ea

Gluing the bulk and defect $\mathbb{V}$ operators at the defect point leads to sewing
conditions that relate the bulk fields at the defect point with the defect fields. In
particular, setting
\be
\mathbb{V}_{(0)}^+(x\to x_0) = \mathbb{\widetilde{V}}_{(0)}^+(x_0)~,
\ee
leads to a unique sewing condition for the four matrix entries
\be
\mathcal{C}^{(0)}: \qquad \xi^+ - a^-a^+ \xi^- -(a^-+a^+)\xi^z  = 0~.
\label{sewing_0}
\ee
Solving this condition
with respect to $\xi^-$ and substituting the solution back to the defect contribution
of the first charge \eqn{deffstch} simplifies the latter one significantly, which can
be written in a very compact form as
\be
\mathcal{D}^{(0)} = a^-\xi^- + \xi^z~.
\ee
The sewing condition emerging when gluing the $\mathbb{V}$ operators at the defect point
from the left is related with eq. (\ref{sewing_0}) by complex conjugation.

At the next order, the modified $\mathbb{V}$ operator at the defect point from the right
is
\[
\mathbb{\widetilde{V}}_{(1)}^+ = \frac{1}{\m^2}\Big(\frac{\fr{d}^+}{\fr{J}}\Big)^2
\begin{pmatrix}
 \Om_{11} & \Om_{12} \cr \Om_{21} & \Om_{22}
\end{pmatrix} ~,
\]
with
\begin{subequations}
\ba
\Om_{11} & =&  \left(a^- \bar{a}^{+'} + (a^{-'} - a^-) \bar{a}^+ \right)\xi^+\xi^-
+ a^-\left(a^- \bar{a}^+ - a^- \bar{a}^{+'} -2\right) \xi^- \xi^z \cr
& + & \left(\bar{a}^{+'} - \bar{a}^+\right)\xi^+\xi^z
+\left(a^-\bar{a}^+ -1 + a'^-\bar{a}^+ - a^- \bar{a}'^+\right)(\xi^z)^2 - (a^-\xi^- )^2
\ea
\ba
\Om_{12} & = & \bar{a}^+(a^{-'} - a^-)\xi^+\xi^- + a^-\left(\bar{a}^+(a^-\bar{a}^+-2)
-2 \bar{a}^{+'}\right) \xi^- \xi^z - (\bar{a}^+)^2\xi^+ \xi^z \cr
& + &  \left(\bar{a}^+(-1+a^-\bar{a}^+ -1 + \bar{a}^+ a^{-'}\right)
(\xi^z)^2 - (\bar{a}^+ + \bar{a}^{+'})(a^-\xi^-)^2
\ea
\ba
\Om_{21} & = & (a^-)^2 \xi^- \xi^z + (2\bar{a}^+ a^- -2a^-\bar{a}^{+'})\xi^+ \xi^z
- (a^- + a^{-'} )\xi^+ \xi^- \cr
& + & \big(a^- - \bar{a}^+ (a^-)^2 + (a^-)^2 \bar{a}^{+'} - a^{-'}\big)(\xi^z)^2
+ (\bar{a}^{+'} - \bar{a}^+)(\xi^+)^2
\ea
\ba
\Om_{22} & = & -\big(a^- \bar{a}^{+'} + \bar{a}^+ (a^- + a^{-'}) \big) \xi^+ \xi^-
+ (a^-)^2 (\bar{a}^+ + \bar{a}^{+'}) \xi^- \xi^z - (\bar{a}^+ \xi^+)^2 \cr
& + & \big(\bar{a}^+ a^- -(\bar{a}^+ a^-)^2 + a^- \bar{a}^{+'} - \bar{a}^+ a^{-'}
\big)(\xi^z)^2  - \big( \bar{a}^+(1 - 2 \bar{a}^+ a^-) -\bar{a}^{+'} \big) \xi^+ \xi^z ~.\cr
&&
\ea
\end{subequations}
Equating the first order Lax operators at the defect point, $\mathbb{V}_{(1)}^+(x\to x_0)
= \widetilde{\mathbb{V}}_{(1)}^+(x_0)$, yields a unique second sewing  condition,
$\mathcal{C}^{(1)}$, for all four entries of the matrices, which is
rather complicated and long. However, using the previously
found sewing condition of the lower order, $\mathcal{C}^{(0)}$, the new condition
simplifies greatly and assumes the form
\[
\bar{a}^+ (\fr{d}^+)^2 (a^- \xi^- + \xi^z) \left(a^- + a^- a^{+'} \xi^- +
   a^+ (-1 + a^{-'} \xi^-) + (a^{-'} + a^{+'}) \xi^z\right) = 0~,
\]
or, in other words,
\be
\mathcal{C}^{(1)}: \qquad
(a^- - a^+) + ( a^- a^+)'\, \xi^-  + (a^{-} + a^{+})'\, \xi^z = 0~,
\label{second_sc}
\ee
since the factor $\bar{a}^+ (\fr{d}^+)^2$ in front cannot be zero and the next term
inside the parenthesis is just $\mathcal{D}^{(0)}$. We observe that the sewing condition
at this order involves not only the left and right fields, as well as the defect fields,
but also the derivatives of the bulk fields (see also \cite{avan-doikou-defect1, avan-doikou-defect2}).
Proceeding to higher orders, one builds an infinite tower of
sewing conditions, which in principle contain higher order derivatives of the bulk and
defect fields. In fact, roughly speaking the sewing conditions of each order are 
associated to the corresponding integral of motion. In the same spirit that there 
exists an infinite tower of independent charges in involution in integrable field 
theories, so that the system is integrable, an infinite tower of independent sewing 
conditions also exists (see also relevant proof on the invariance of the sewing 
conditions under the generic Hamiltonian action in \cite{avan-doikou-defect1}). In 
any case, this is not really surprising, given for instance that the same scenario 
applies in integrable field theories with non trivial boundaries, where an infinite tower 
of suitable boundary conditions is also consistently extracted \cite{avan-doikou-boundary}.
Moreover, given that the sewing conditions reflect the
analyticity of the fields at the defect point, we expect that they should provide useful
information in order to solve the equations of motion derived below.

It has been explicitly checked that the extracted sewing conditions for
the first two orders, i.e. (\ref{sewing_0}) and (\ref{second_sc}),  are as expected
compatible with the Hamiltonian hierarchy, confirming the relevant generic statement
proven in \cite{avan-doikou-defect1} regarding the invariance of all sewing
conditions under the Hamiltonian action. Note also that the commutativity of the
momentum and  Hamiltonian
is valid as expected, but {\it on-shell}, i.e. the sewing conditions must be taken into
account, as it happens in the case of the sine-Gordon model with defect (see also
\cite{avan-doikou-defect2}),
\be
\Big \{{\cal H},\ {\cal P} \Big \} =0\, .
\ee
It is worth noting that taking into account the second sewing condition
(\ref{second_sc}), the first order defect contribution simplifies significantly;
it is eventually given by the following simple expression
\be
\mathcal{D}^{(1)} = \frac{a^{+'} + a^{-'}}{a^+ - a^-}~.
\ee
It is interesting that on-shell the defect fields $\xi^i$ cancel out completely from
the defect Hamiltonian. This simplification could be a hint that a simpler formulation
of implementing dynamical defects in the theory could exist. On the other hand, it
is possible that this effect emerges as a bi-product from the existence of integrability
of the model. Furthermore, one should keep in mind that this effect is present only
in the on-shell Hamiltonian of the model, once the sewing conditions are taken into
account, and is related to the on-shell versus off-shell integrability issue of the
model. For a more detailed discussion on the on-shell vs off shell integrability we
refer the interested reader to \cite{avan-doikou-defect2}.

Finally, the equations of motion for the left and right bulk theories are given by (\ref{emh}),
for the right theory the variables are simply $\tilde S_i$. The time evolution on the defect point is expressed as:
\ba
&& \dot \xi^- = {2 \over \fr{J}} \Big ( {\mathrm A} - {\cal D}^{(1)}\bar a^+ \fr{d}^+\Big )\xi^z + {1 \over \fr{J}} \Big ({\cal D}^{(1)}\fr{d}^+(1-a^-\bar a^+) -{\mathrm C}\Big )\xi^- \cr
&& \dot \xi^+ = {2 \over \fr{J}} \Big ( -{\mathrm B} + {\cal D}^{(1)}a^- \fr{d}^+\Big )\xi^z + {1\over \fr{J}} \Big (-{\cal D}^{(1)}\fr{d}^+(1-a^-\bar a^+) +{\mathrm C}\Big )\xi^+  \cr
&& \dot \xi^z = -{1 \over \fr{J}} \Big ( {\mathrm A} - {\cal D}^{(1)}\bar a^+ \fr{d}^+\Big )\xi^+ +
{1 \over \fr{J}} \Big ({\mathrm B} -{\cal D}^{(1)}a^- \fr{d}^+\Big )\xi^-\, ,
\ea
where we define
\ba
&&{\mathrm A} = \bar a^{+'} + (\bar a^+)^2 a^{+'},
~~~~~{\mathrm B} = (a^+ - a^{+'})a^-\bar a^{+'} + \fr{d}^+ a^{-'}, \cr &&
{\mathrm C}=
\bar{a}^{+'} (a^+ + a^-) + \bar{a}^+ (\bar{a}^+ a^{+'} a^- -a^{+'}  -\fr{d}^+ a^{-'})\, .
\ea
The equations above, relevant to the defect point, emerge either from the Hamiltonian description through
\be
\dot \xi^i =\Big \{{\cal H},\ \xi^i \Big \}\, ,
\ee
or the zero curvature condition on the defect point (\ref{zerod}). Recall that
throughout this work we consider only dynamical type defects, so that
the equations of motion above describe their time evolution. It is an interesting task
to attemp to solve this set of equations, as well as determining the nature of the
physical information obtained from the solutions. It should be stressed out that
not only the sewing conditions presented before should have to be taken into account 
when solving the equations of motion, but it is expected that they will also simplify 
the somehow complicated expressions of the latter ones. We hope to address these
interesting issues in a forthcoming publication, starting from the less complicated
NLS defect model \cite{avan-doikou-defect1}.

\section{The Faddeev-Reshetikhin model with defect}
The F-R model was introduced in \cite{Faddeev:1985qu}, as a convenient modification
of the familiar non-ultra local PCM model. Despite the different underlying algebras,
it was shown that the equations of motion of these two models coincide, hence they may be considered
as equivalent. Moreover, the F-R model is easier to quantize, due to the
absence of non-ultra local terms in the algebra.

\subsection{The model}
The Lax operator of the model is the same with the familiar PCM one and has the form
\be
{\mathbb U}(x, \lambda)= \frac{\l~J_0(x) +J_1(x)}{\l^2 -1}\, ,
\ee
with $J_i$ being currents taking values on the Lie algebra $\fr{g}$, corresponding
to a Lie group $G$. In the present work, we restrict ourselves to the simple case of the
$SU(2)$ group. We choose the Pauli matrices $(\s^+, \s^-, \frac{1}{2}\s^z)$ as the generators of the
$\fr{su}_2$ Lie algebra. We also define the currents
\be
S = J^+ = \frac{J_0 + J_1}{2}, \qquad T = J^- = \frac{J_0 - J_1}{2}\, ,
\ee
so that the Lax operator may also be written as
\be
{\mathbb U}(x,\lambda) = \frac{S}{\l -1} + \frac{T}{\l+1}\, .
\ee
Expanding in the Lie algebra generators, the currents are expressed in the following component
form
\be
S = S^a t^a = S^z \tfrac{\s^z}{2} + S^+ \s^+ + S^- \s^-\, ,
\ee
and similarly for $T$. The time component of the Lax pair is given by:
\be
{\mathbb V}(x, \lambda) = -\frac{S}{\l -1} + \frac{T}{\l+1}\, .
\label{bulkv}
\ee

The Lax operator satisfies the fundamental Poisson bracket relation
\be
\poisbrac{{\mathbb U}(x,\l)}{{\mathbb U}(y,\m) } = \Big [r(\l -\m), ~{\mathbb U}(x,\l)\otimes \mathbb{I} +
\mathbb{I} \otimes {\mathbb U}(x,\m) \Big ] \d(x-y)\, ,
\ee
which reflects the absence of non-ultra-local terms in the level of the algebra for
$S$ and $T$. In particular, the latter ones satisfy the following algebraic relations
\ba
\Big \{ S^a(x) , ~S^b(y)\Big  \} & = & f^{ab}_{~~c}~S^c(x) \d(x-y)\cr
\Big \{ S^a(x) , ~T^b(y)\Big  \} & = & 0\cr
\Big \{ T^a(x) , ~T^b(y)\Big  \} & = & f^{ab}_{~~c}~	T^c(x) \d(x-y)\, .
\ea
As mentioned above, we restrict our attention in the simplest case, i.e. the $SU(2)$
group. The underlying algebra consists of two independent copies of the $\fr{su}_2$
Lie algebra and the structure constants are proportional to the usual Levi-Civita tensor
($f^{ab}_{~~c} = 2 i \varepsilon_{abc}$). Moreover, $S$ and $T$ may be interpreted as
three-vectors and their lengths can be fixed as: $|S| = |T| = \mathcal{C}= constant$;
i.e. they are simply the $\fr{su}_2$ Casimir operators. Note that the latter restriction
($\mathcal{C}= constant$) is known as the Virasoro constraint in the string theory frame
(see e.g. \cite{kaza}), rendering essentially the familiar PCM model ultra-local.

\subsection{Local integrals of motion}
Let us first briefly review the bulk theory. From the Lax operator of the F-R model, it is
manifest that there are two expansions that one may obtain, around two singular
points. The local conserved charges, such as the momentum and the Hamiltonian are then
linear combinations of the quantities derived. We present details for the $\l=1,\ ({\mathrm y}=\l-1=0)$ expansion,
the results for the $\lambda =-1$ are similar and are omitted. In particular, for the
bulk parts we have
\be
W = \sum_{n=0}^{\infty} {\mathrm y}^n W_n= \sum_{n=0}^{\infty}{\mathrm y}^n
\begin{pmatrix}
 0 & a_n \cr b_n & 0
\end{pmatrix},
\ee
and
\be
Z = \sum_{n=-1}^{\infty}{\mathrm y}^n Z_n= \sum_{n=-1}^{\infty}{\mathrm y}^n
\begin{pmatrix}
  \g_n & 0 \cr 0 & \d_n
\end{pmatrix}.
\ee
For the first order we have found for $W$
\be
a_0 = \frac{1}{2S^+}(S^z \mp \mathcal{C}), \qquad -\bar{a}_0 =
b_0 = \frac{1}{2S^-}(-S^z \pm \mathcal{C})\, ,
\label{1stordFR}
\ee
with
\be
\mathcal{C} \equiv \sqrt{(S^z)^2 + 4(S^-)(S^+)} = 2\textrm{tr}(S^2) = 2\textrm{tr}(T^2)
=\sqrt{(T^z)^2 + 4(T^-)(T^+)}\, .
\ee
Setting the Casimir $\mathcal{C}$ equal to one amounts to coinciding with the results
of the L-L model. In the next order
\ba
\pm a_1~(2\mathcal{C}) & = & -T^- - a_0~T^z + a_0^2~T^+ + 2a_0' \cr
\pm ~b_1~(2\mathcal{C}) & =& T^+ - b_0~T^z - b_0^2~T^- - 2b_0'~.
\ea
Observe also that
\be
b_1 = -\bar{a}_1 + \tfrac{2}{\mathcal{C}}\, \bar{a}_0'~,
\ee
so that everything can be written in terms of $a_0$ and $a_1$, which may now be regarded
as the fundamental fields.
We carry on the computations for the $Z$ elements. The results we have found are the
following:
\ba
2~\g_{-1}' & = & + S^z + 2b_0~S^- = \pm\mathcal{C}\cr
2~\d_{-1}'&  = & -S^z + 2a_0~S^+ = \mp\mathcal{C}\cr
 4~\g_0'&  = & +T^z + 2b_0~T^- + 4b_1~S^- \cr
4~\d_0' &  = & - T^z + 2a_0~T^+ + 4a_1~S^+~.
\ea
It is manifest that the first integrals are trivial since the corresponding densities
are constant
\be
\mathcal{I}_{-1} = \mathcal{\hat{I}}_{-1} = L\, \mathcal{C}~.
\ee

Henceforth we choose to consider ${\cal C}=1$, as well as the first of the two signs appearing
in all the expressions above.
Let us first derive the expression for the bulk theory with Schwartz type boundary conditions.
Substituting the exact expressions for $a_i$'s and $b_i$'s we obtain
the charges in terms of the fields. Bear in mind that some boundary terms
vanish due to Schwartz boundary conditions, that will have to be taken into account
when considering non-trivial boundary conditions at the defect point. Eventually, one is
left with
\ba
{\cal I}_0 & = & \frac{1}{2} \mathcal{P}_S
- \frac{1}{2} \int_{-L}^L dx\,  \textrm{tr} (S~T) \cr
\hat {\cal I}_0 & = & \frac{1}{2} \mathcal{P}_T
+ \frac{1}{2} \int_{-L}^L dx\, \textrm{tr}  (S~T)~,
\ea
and
\be
\mathcal{P}_S = i \int_{-L}^{L}dx\  \frac{S_1S'_2 - S'_1 S_2}{1 + S^z}, \qquad
\mathcal{P}_T = i \int_{-L}^{L}dx \frac{T_1T'_2 - T'_1 T_2}{1 + T^z}\, .
\ee
Taking their sum and their difference yield the Hamiltonian and momentum densities of the
system respectively. More specifically,
\ba
\mathcal{P} & = & {1\over 2}\mathcal{P}_S + {1\over 2}\mathcal{P}_T \cr
\mathcal{H} & = & {1\over 2}\mathcal{P}_S - {1\over 2}\mathcal{P}_T
- \int_{-L}^L dx\,\textrm{tr}  (S~T)~.
\ea
The above results hold for the right and left bulk parts of the F-R model. Recalling the
generic expression for the generating function \eqn{genfundef}, we find the defect
contributions at all orders.

Now consider the system in the presence of a point-like integrable defect at the position
$x_0$, described by the generic $L$-matrix (\ref{defectmatrix}).
The expressions of the momentum and the Hamiltonian emerging from the expansion of $\lambda$
around the values $\pm 1$, are given as:
\ba
{\cal I}_0 &=& {1 \over 2} \int_{-L}^{x_0} dx\ \Big (i{S_1S'_2 -S'_1 S_2 \over 1 +S_3}
- \textrm{tr}(ST) \Big )+  {1 \over 2} \int_{x_0}^{L} \Big (i {\tilde S_1\tilde S'_2 -\tilde S'_1 \tilde S_2 \over 1 +\tilde S_3}
-  \textrm{tr}(\tilde S \tilde T) \Big ) \cr
&-& {1\over 2} \ln \Big ((1+S_3)(1 +\tilde S_3)\Big)(x_0) - \ln\fr{V}(x_0)\, ,
\ea
\ba
\hat {\cal I}_0 &=& {1 \over 2} \int_{-L}^{x_0} dx\ \Big (i{T_1T'_2 -T'_1 T_2 \over 1 +T_3}
+  \textrm{tr}(ST) \Big )+  {1 \over 2} \int_{x_0}^{L} \Big (i {\tilde T_1\tilde T'_2 -\tilde T'_1 \tilde T_2 \over 1 +\tilde T_3}
+ \textrm{tr}(\tilde S \tilde T) \Big ) \cr
&-& {1\over 2} \ln \Big ((1+T_3)(1 +\tilde T_3)\Big)(x_0) - \ln \fr{\hat V}(x_0)\, ,
\ea
where we define
\ba
\fr{V} &=& \xi^z(1+a_0^+b_0^-) -a_0^+\xi^+ + b_0^-\xi^- + 1 - a_0^+ b_0^- \cr
\fr{\hat V} &=&  \xi^z(1+\hat a_0^+\hat b_0^-) -\hat a_0^+\xi^+ + \hat b_0^-\xi^- - 1 +\hat a_0^+ \hat b_0^- \, ,
\label{frv}
\ea
where $a_0^-,\ b_0^-$ are defined as in (\ref{1stordFR}) with the $S$ fields,
$a_0^+,\ b_0^+$ are defined with the $\tilde S$ fields, and
\be
\hat a_0^- = {T_3-1 \over 2T^+}, ~~~~\hat b_0^- = {1-T_3 \over 2T^-},
~~~~\hat a_0^+ = {\tilde T_3-1 \over 2\tilde T^+},
~~~~\hat b_0^+ = {1-\tilde T_3 \over 2\tilde T^-}\, .
\ee
The fields $S,\ T$ correspond to the left theory whereas the fields $\tilde S,\ \tilde T$ correspond to the right theory.
The complete momentum and Hamiltonian of the system are given by the following expressions:
\ba
{\cal P} &=& {\cal I}_0 + \hat {\cal I}_0 \cr
{\cal H} &=&  {\cal I}_0 - \hat {\cal I}_0\, .
\ea

\subsection{The modified Lax pair}
The next step is the derivation of the associated time components of the Lax pairs for the
bulk theories and at the defect point, from left and right. Requiring continuity around
the defect point the relevant sewing conditions are extracted.
The bulk quantity associated to the Hamiltonian is given by (\ref{bulkv}). The
$\widetilde {\mathbb V}$ operator at the defect point from the right, from the expansion
around $\lambda=1$ has the following expression:

\begin{subequations}
\be
\widetilde {\mathbb V}_0^+ = -{1 \over (\lambda-1)\fr{V}}
\begin{pmatrix}
 \xi^z+1 +b_0^-\xi^- & -a_0^+(\xi^z+1) -a_0^+b_0^- \xi^- \cr
 \xi^+ + b_0^- (-\xi^z+1) & -a_0^+\xi^+ -a_0^+b_0^-(-\xi^z+1)
\end{pmatrix}\, ,
\ee
while from the expansion around $\lambda=-1$ one obtains
\be
\hat{\widetilde {\mathbb V}}_0^{+} = -{1 \over (\lambda+1) \fr{\hat V}}
\begin{pmatrix}
 \xi^z-1 +\hat b_0^-\xi^- & -\hat a_0^+(\xi^z-1) -\hat a_0^+\hat b_0^- \xi^- \cr
 \xi^+ + \hat b_0^- (-\xi^z-1) & -\hat a_0^+\xi^+ -\hat a_0^+\hat b_0^-(-\xi^z-1)
\end{pmatrix}\, ,
\ee
where $\fr{V}$, $~\fr{\hat V}$ are defined in (\ref{frv}).
Similar expressions hold for $\widetilde {\mathbb V}^-$, in particular:
\be
\widetilde {\mathbb V}_0^- = -{1 \over (\lambda-1)\fr{V}}
\begin{pmatrix}
 \xi^z+1 -a_0^+\xi^+ & a_0^+(\xi^z-1) +\xi^- \cr
 -a_0^+b_0^-\xi^+ + b_0^- (\xi^z+1) & b_0^-\xi^- -a_0^+b_0^-(-\xi^z+1)
\end{pmatrix}\, ,
\ee
and
\be
\hat{\widetilde {\mathbb V}}_0^- = -{1 \over (\lambda+1)\fr{\hat V}}
\begin{pmatrix}
 \xi^z - 1 -\hat a_0^+\xi^+ & \hat a_0^+(\xi^z + 1) +\xi^- \cr
 -\hat a_0^+\hat b_0^-\xi^+ + \hat b_0^- (\xi^z - 1) &\hat b_0^-\xi^- +\hat a_0^+\hat b_0^-(\xi^z + 1)
\end{pmatrix}\, .
\ee
\end{subequations}
Due to the continuity
requirements, the corresponding sewing conditions arise, from the expansion around $\lambda =1$:
\be
{\cal C}^{(0)}: ~~~~~~~\xi^+ - b_0^+b_0^- \xi^- - (b_0^+ + b_0^-) \xi^z - (b_0^+-b_0^-)=0 \, ,
\ee
and from the expansion around $\lambda =-1$:
\be
\hat {\cal C}^{(0)}: ~~~~~~~\xi^+ - \hat b_0^+\hat b_0^- \xi^- - (\hat b_0^+ + \hat b_0^-) \xi^z
- (\hat b_0^+-\hat b_0^-)=0\, .
\ee
Commutativity between the momentum and the Hamiltonian of the F-R system is shown via
the algebraic relations of the defect degrees of freedom, and as expected:
\be
\Big \{ {\cal H},\ {\cal P} \Big \}=0\, .
\ee
Furthermore, as in the case of the L-L model, the compatibility of the sewing constraints
with the charges in involution is explicitly checked and confirmed.

Finally, by means of the underlying Poisson structure, or the zero curvature condition,
one may derive the equations of motion for the bulk theories
\ba
&& \dot S(x, t) =-S'(x, t)  -\Big [ S(x, t),\ T(x, t) \Big ] \cr
&& \dot T(x, t)  = T'(x, t)  -\Big [T(x, t),\ S(x, t) \Big ], ~~~~~~~x \neq x_0\, .
\ea
Similar equations of motion hold for the fields of the right theory $\tilde S,\ \tilde T$,
which are omitted here for brevity. The time evolution of the defect degrees of freedom
reads as:
\ba
&& \dot \xi^+ = 2 \Big ({{\mathrm b} \over \fr{V}}  - {\hat {\mathrm b} \over \fr{\hat V}}\Big )\xi^z -
\Big ( {{\mathrm c}\over \fr{V}}  - {\hat {\mathrm c} \over \fr{\hat V}}\Big )\xi^+\cr
&& \dot \xi^- = -2 \Big ({{\mathrm a} \over \fr{V}}  - {\hat {\mathrm a} \over \fr{\hat V}}\Big )\xi^z +
\Big ({{\mathrm c} \over \fr{V}}  - {\hat {\mathrm c} \over \fr{\hat V}}\Big )\xi^- \cr
&& \dot \xi^z =\Big ({{\mathrm a} \over \fr{V}}  - {\hat {\mathrm a} \over \fr{\hat V}}\Big )\xi^+ -
\Big ({{\mathrm b} \over \fr{V}}  - {\hat {\mathrm b} \over \fr{\hat V}}\Big )\xi^- \,.
\label{evolution}
\ea
where we define
\ba
&& {\mathrm a}=-a_0^+, ~~~~~{\mathrm b}=b_0^-, ~~~~~{\mathrm c} = 1 +a_0^+b_0^- \cr
&&\hat {\mathrm a} = -\hat a_0^+,~~~~~\hat {\mathrm b}=\hat b_0^-, ~~~~~\hat {\mathrm c} = 1 +\hat a_0^+ \hat b_0^-.
\ea
Note that expressions (\ref{evolution}) arise either from the Hamiltonian or from the zero
curvature condition at the defect point. With this, we complete our analysis on sigma
models in the presence of dynamical point-like defect. In the last section we summarize our
results and discuss some interesting extensions that will be analyzed in future works.

\section{Discussion}

We have applied the methodology developed in \cite{avan-doikou-defect1} for the
Landau-Lifshitz and the Faddeev-Reshetikhin models. Due to the existence of the
classical $r$-matrix, we have been able to formulate a modified monodromy matrix, and
hence identify the hierarchy of Poisson commuting Hamiltonian alongside the
associated Lax pairs. Certain sewing conditions compatible with the Hamiltonian action
have also been identified.

In the present and previous studies dynamical type defects have been considered, the so
called type-II defect according to the terminology of \cite{BCZ1, BCZ2}. This type of
defects is associated to a Poisson structure strictly related to the point-like defect,
hence time evolution of the defect degrees of freedom occurs. The next natural step would
be the investigation of the type-I or non-dynamical defects. Study of non-dynamical defects
may lead to B\"acklund transformations associated to the defect point as pointed out
in \cite{BCZ1, BCZ2}. It would be quite interesting to check whether these B\"acklund
transformations could naturally emerge through the proposed algebraic formulation.
For instance if the defect is movable then possibly there exists some B\"acklund
transformation associated, although this is not a priori clear. Nevertheless, still
the question of the relevant Poisson structure is raised. Another key issue raised
is about dressing and its compatibility with the Poisson structure, in other words
is the dressing/B\"acklund transformation group a Lie Poisson group? This is true
usually in the bulk case, but not obvious in the defect case.
Regarding the dynamical defects that we have studied here, despite our efforts, we
 have not been able so far to show the natural emergence of a similar geometric origin, which
consequently remains an open issue even in the frame of other models studied in earlier
works \cite{avan-doikou-defect1, avan-doikou-defect2}.

We have restricted our attention so far to point-like defects, although in principle
spatially extended impurities may also be investigated within the proposed scheme.
Moreover, at the quantum level one may study the associated Bethe ansatz equations, and
hence extract significant information regarding the interaction of particle like excitations
displayed by such models with the defect. All the above are highly intricate issues and
hopefully will be addressed in forthcoming publications

\paragraph{Acknowledgements\\}
NK acknowledges financial support provided by the Research Committee of the University of Patras
via a K. Karatheodori fellowship, under contract number C.915. He would also like to thank the
Physics Division of the National Technical University of Athens for kind hospitality during
the completion of this work.

\end{document}